\documentclass[runningheads]{llncs}
\usepackage[T1]{fontenc}
\usepackage{graphicx}
\usepackage[english]{babel}\addto\captionsenglish  
{}  
\usepackage{blindtext}
\usepackage[sorting=none]{biblatex}
\usepackage{tabularx}
\usepackage{biblatex}  
\usepackage[utf8]{inputenc}
\usepackage[caption=false]{subfig}
\usepackage{graphicx}
\usepackage{mathtools}
\usepackage{pifont}
\usepackage{comment}
\usepackage{algorithm}
\usepackage{mathtools}
\usepackage{authblk}
\usepackage{hyperref} 
\usepackage{amsmath}
\usepackage{algorithm}
\usepackage{algorithmic}
\usepackage{tabulary} 
\usepackage{booktabs} 
\usepackage{multirow} 
\usepackage{array} 
\usepackage{caption}
\begin{document}

  \title{SecFlow: Adaptive Security-Aware Workflow Management System in Multi-Cloud Environments}

\titlerunning{SecFlow: Adaptive Security-Aware Workflow Management System}
%
\author{Nafiseh Soveizi\inst{1}\orcidID{0000-0003-2111-734X} \and
Fatih Turkmen\inst{1}\orcidID{0000-0002-6262-4869}}
\authorrunning{N.Soveizi and F.Turkmen}
%
\institute{Information Systems Group, University of Groningen, Groningen, The Netherlands
\email{\{n.soveizi,f.turkmen\}@rug.nl}}
\maketitle              
\begin{abstract}
In this paper, we propose an architecture for a security-aware workflow management system (WfMS) we call SecFlow in answer to the recent developments of combining workflow management systems with Cloud environments and the still lacking abilities of such systems to ensure the security and privacy of cloud-based workflows. The SecFlow architecture focuses on full workflow life cycle coverage as, in addition to the existing approaches to design security-aware processes, there is a need to fill in the gap of maintaining security properties of workflows during their execution phase. To address this gap, we derive the requirements for such a security-aware WfMS and design a system architecture that meets these requirements. SecFlow integrates key functional components such as secure model construction, security-aware service selection, security violation detection, and adaptive response mechanisms while considering all potential malicious parties in multi-tenant and cloud-based WfMS.

\end{abstract}
    \section{Introduction}
\label{sec:intro}
 In recent years, workflows are the commonly used application model to describe both business and scientific workflows. A workflow defines a series of computational tasks logically connected by data- and control-flow dependencies \cite{Dumas2018}. Using workflows to specify complex processes makes the management of such processes easier and more consistent in a structured, distributed, and automated manner. Workflows are managed by Workflow Management Systems (WfMSs) that are responsible for receiving the workflow input from its users and producing the output of each workflow execution (a.k.a. instance), and at the same time providing essential functionality to enable the execution of workflows such as task scheduling, service composition, managing the data- and control-flow dependencies, resource provisioning, and fault tolerance 
 \cite{Li2012a, Rodriguez2017}. 
 In addition to workflow modeling and execution, WfMSs play a crucial role in managing and analyzing operational processes. They serve as essential tools for enhancing effectiveness, efficiency, cost-effectiveness, quality, and productivity improvements. \cite{meidan2017survey,poola2017taxonomy}. 

The recent developments towards supporting data- and compute-intensive applications led to the need for flexible and scalable workflows and WfMS to fulfill users' requirements. Cloud-based WfMS is an effective solution providing the ability for the system to scale up or down resources as needed to meet changing demands. Besides, cloud-based solutions can also compensate for the limited processing capabilities of the users by outsourcing all or part of client-side operations to the cloud. 

Cloud security significantly impacts the utilization of cloud services and infrastructures, particularly for workflows involving sensitive data and tasks~\cite{SOVEIZI2023, Varshney2019}. In fact, when a workflow or part of it is outsourced to the cloud, it will lead to increased security risks and make them vulnerable to malicious attacks. 
Deploying the entire WfMS on a semi-trusted or untrusted cloud further exacerbates the situation.
Consequently, the security properties of workflows are inevitably affected. Therefore, it is crucial to identify potential malicious entities and other security threats within such systems and establish a secure architecture that efficiently addresses these risks through effective security mechanisms. This conclusion is based on the findings of a recent literature review of the security and privacy concerns in both scientific and business workflows~\cite{SOVEIZI2023} in which we investigated the current state of the art and its limitations.
Our findings show that currently available research does not address security throughout the entire workflow lifecycle although it is essential in order to prevent cascading effects and the increased difficulty and cost associated with detecting and containing security issues in later phases. Furthermore, we could conclude that there is a widely unexplored area of research connected to detecting, predicting, and reacting to security violations during the execution time of cloud-based workflows.

To bridge this gap in the literature and tackle the challenges mentioned, our paper introduces SecFlow, a security-aware WfMS designed to address the essential requirements of such a system, with a primary focus on security and privacy. Towards this goal, the contributions of this paper are summarized as follows:

\begin{itemize}
\item We provide a classification of the possible security attacks on cloud-based workflows in our multi-tenant WfMS and thus establish the security requirements for a secure WfMS.
\item We propose a security-aware functional architecture for a WfMS that meets the identified requirements and evaluate its performance.
\end{itemize}

By addressing the identified gaps in the state-of-the-art~\cite{SOVEIZI2023}, our contribution focuses on developing a WfMS that effectively mitigates security risks in a multi-cloud environment. SecFlow provides comprehensive security measures throughout the entire workflow lifecycle and considers all potential malicious parties, thereby addressing a critical need in the field.

Our paper has the following structure: Firstly, in section \ref{sec:requirements}, we will present the classification of the security vulnerabilities of a cloud-based WfMS and then discuss the potential countermeasures against them known from the literature. After that, the existing cloud-based and security-aware WfMSs are discussed in section \ref{relatedwork}. In section~\ref{sec:architecture}, the proposed architecture is described in detail. We evaluate the proposed architecture in section~\ref{evaluation}. Finally, section~\ref{conclusion} presents our conclusions. 
    \section{Security in Cloud-based Multi-tenant WfMSs}
\label{sec:requirements}
In this section, we examine the security vulnerabilities of a WfMS, following the Open Web Application Security Project (OWASP) terminology \cite{OWASPThreatModeling}. We begin by emphasizing the importance of sensitive resources, referred to as \textit{Assets}, in multi-tenant WfMS and explore potential threats from attackers, known as \textit{Actors}, within these environments.
Furthermore, we analyze the impact of various attacks on the respective targets and explore the preventive measures, referred to as \textit{Preventions} as well as the \textit{Mitigations} commonly employed to minimize the impact of such attacks. To provide a clear overview of these security concerns, preventions, and mitigations, we present Figure \ref{fig:attackClassification} as a concise visual representation of this classification.

\subsection{Assets of Tenants}
\label{sec:Assets}
In multi-tenant WfMSs, the assets of tenants, which consist of sensitive resources, are the primary targets for potential attacks. This section discusses three valuable assets owned by each tenant.

The \textbf{Tenant's Metadata} includes sensitive data such as account information and the number of users. Unauthorized disclosure of this information can compromise the overall security of the tenant.

Each tenant has different users who need to undergo the authentication process to access tasks and resources. Compromising \textbf{Users' Metadata}, such as legitimate user account credentials, can lead to violations of Confidentiality, Integrity, and Availability (CIA) of the users' tasks.

The \textbf{Workflow} is the most significant asset in the WfMS and needs to be Protected at different levels of abstraction. It encompasses the following assets:

\textit{\textbf{Tasks}}: Workflows consist of various types of tasks, including user tasks and service tasks. These tasks can be performed by users or outsourced to the cloud.

\textit{\textbf{Intermediate data}}: Another critical asset of workflows is the intermediate data generated by the tasks, which includes the data that the workflow exchanges with external services and users via a network.

\textit{\textbf{Logic}}: Workflow logic represents another important asset that should be protected from reconstruction and disclosure by third parties.

\subsection{Potential Actors}
\label{sec:PotentialActors}

\begin{figure}[bth!]
\includegraphics[width=\columnwidth]{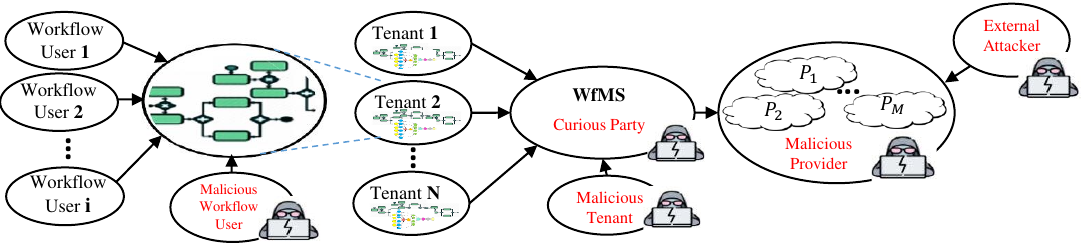}
\caption{Adversary model in cloud-based and multi-tenant WfMS}
\label{fig:potential actors}
\end{figure}

In multi-tenant and cloud-based WfMS, various entities have the potential to compromise the assets of tenants at different phases of the workflow lifecycle. This section discusses the details of these potential actors. Figure \ref{fig:potential actors} illustrates our adversary model within the system.

\textbf{Workflow users} encompass all parties involved in the workflow tasks, including roles, organizational units, or the entire organization. It is important to acknowledge that these users have the capability to engage in both intentional and unintentional malicious activities.

In a multi-tenant environment, \textbf{tenants} themselves have the potential to become threat actors. Malicious tenants can launch various attacks to compromise the assets of other tenants. Their objectives may involve unauthorized access to sensitive information, manipulation of tasks and data, or causing disruptions.

The cloud-based \textbf{WfMS} is responsible for providing essential functionality to manage the execution of the workflows which are submitted by tenants. While it is assumed to be semi-trusted and compliant with protocols, it may attempt to gather as much information as possible about tenants and their sensitive data.

\textbf{Cloud providers}, whose infrastructure and services are utilized by the WfMS for executing workflow tasks, can also be considered potential threat actors as semi-trusted parties. They have the ability to exploit tenants' assets, thus posing a significant security risk.

The shared nature of cloud infrastructure, where multiple users utilize computing and storage resources, introduces vulnerability to attacks. \textbf{External Attackers}, including malicious cloud provider users or individuals outside the cloud network (e.g., network attackers), exploit the Internet to execute disruptive attacks, thereby impacting the services available to legitimate users. In our specific case, these actors can target tenant assets during their execution within the cloud providers or even during data transfer processes.

\begin{figure*}[h]
\setlength{\belowcaptionskip}{-10pt}
  \includegraphics[width=\textwidth]{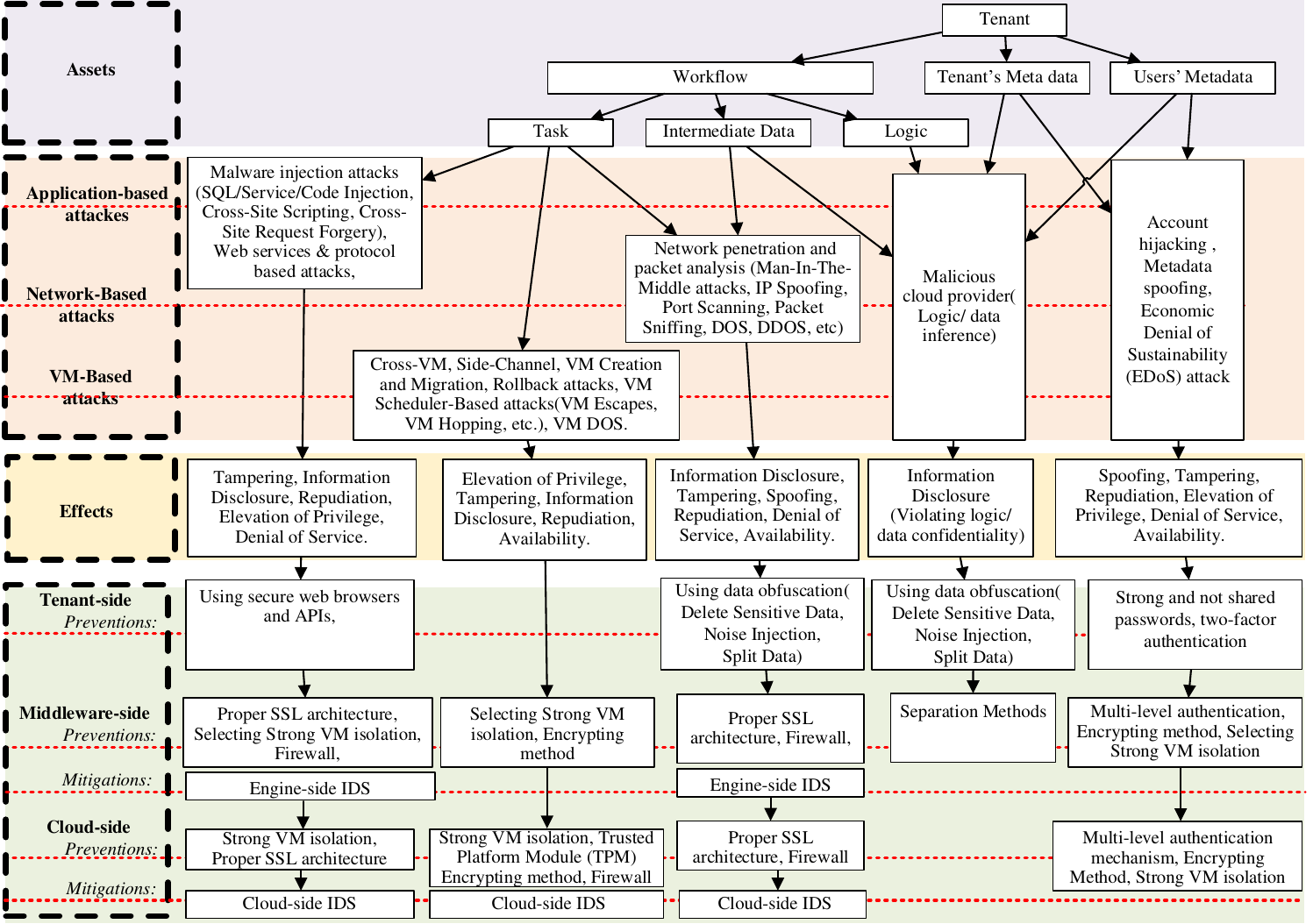}
  \caption{A classification of the possible attacks in the cloud-based and multi-tenant WfMS organized according to assets, levels, effects, and prevention and mitigation approaches.}
  \label{fig:attackClassification}
\end{figure*}
\vspace{-1em} 

\subsection{Attacks and Countermeasures}
This section categorizes potential attacks in cloud-based WfMS into three levels and examines the two existing groups of countermeasures that focus on prevention and mitigation controls against these attacks.

The potential categories of attacks in cloud-based WfMS \cite{Khan2016b} (see \ref{fig:attackClassification}) are: a) Application-based attacks: The applications running on the cloud, such as the engine and tasks, are vulnerable to various attacks, including malware injection and protocol vulnerabilities. b) Network-based attacks: The internal network (a virtual private network) connecting cloud machines and the external network (internet) connecting the cloud front-end to users can be compromised. These attacks can violate the CIA of tasks and data. c) VM-based attacks: These attacks exploit vulnerabilities in Virtual Machines (VMs), compromising the CIA of tasks and affecting cloud services.

In what follows, we discuss the possible attacks as well as the possible prevention and mitigation approaches.


Malware injection attacks, including SQL/Service injection, Cross-site scripting (XSS), and Cross-site request forgery (CSRF), as well as attacks at the level of services or protocols are common threats that significantly impact the CIA of tasks. These attacks can also manipulate the control flow of victim processes. To mitigate these risks, clients should utilize secure interfaces/APIs when interacting with cloud systems. Furthermore, employing a secure implementation of protocols, robust encryption mechanisms, and HW-assisted Trusted Computing methods are crucial for improving security \cite{Khan2016b}. Implementing a web application firewall (WAF) can effectively address common web application-related attacks such as XSS and SQL injection \cite{Modi2017}. Techniques for input validation and neutralization should be applied to sanitize user input \cite{Alhenaki2019}. Additionally, it is essential to have application-based intrusion detection systems (IDS) deployed on both the cloud and middleware side to detect abnormal user activities.

Network penetration and packet analysis are potential attacks that compromise the execution results of tasks, leading to eavesdropping, data leakage, and the unauthorized alteration of task contents before they are passed to the next task. To enhance security in such scenarios, employing secure socket methods like the SSL (Secure Sockets Layer) protocol \cite{Hwang2013} is recommended to ensure secure electronic transactions. Additionally, implementation of firewalls, such as packet-filtering, stateful and proxy firewalls, can effectively detect and prevent unauthorized access to sensitive intermediate data \cite{Maroua2018,Modi2017}. Furthermore, the application of machine learning techniques for detecting anomalous traffic \cite{salman2017machine} on both the cloud side and middleware side can provide further security enhancements.

Logic/data inference by a malicious cloud provider is another possible attack. It can occur when a dishonest provider administrator or high-privileged malicious cloud software combines knowledge of workflow logic to infer sensitive information. This knowledge can be collected by combining data from different workflow tasks or fragments, or by combining data from different workflow instances belonging to the same user/tenant. One solution to mitigate these attacks is splitting sensitive information among different clouds as a prevention control. 

Account hijacking and metadata spoofing present significant threats to the sensitive data of users and tenants. These attacks can compromise the CIA of user and tenant responsibilities. Another attack to be aware of is Economic Denial of Sustainability (EDoS), which targets customers' economic resources by fraudulent billing for resource consumption \cite{bhardwaj2021distributed}. These attacks can manifest at various levels, including the application, network, and VM levels. To mitigate these threats, prevention controls such as utilizing strong and unique passwords, implementing multi-level authentication mechanisms, employing encryption methods, and ensuring robust VM isolation \cite{Alhenaki2019,panda2021survey} can help.

    \section{Related works}
\label{relatedwork}
In this section, we briefly present the existing cloud-based WfMSs that possess some kind of security awareness and how they can handle security requirements during different phases of the workflow lifecycle. More detailed discussion of these systems is available in~\cite{SOVEIZI2023}.

 \begin{table*}[h!]
\fontsize{8pt}{10pt}\selectfont
\caption{{Different WfMSs regarding security concerns in the cloud.} }
\label{table1}
\def\arraystretch{1}
\ignorespaces 
\newcolumntype{L}[1]{>{\raggedright\let\newline\\\arraybackslash\hspace{0pt}}m{#1}}

\begin{tabulary}{\linewidth}{ L{3cm}L{1.7cm}L{1.7cm}L{1.7cm}L{1.8cm}L{1.7cm}}

 \textbf{Feature} &
\textbf{\cite{Wang2018}, 2018}  & 
\textbf{ \cite{Mofrad2019}, 2019}  &
\textbf{ \cite{Kim2015}, 2015} &
\textbf{ \cite{Lins2016}, 2016}&
\textbf{\cite{Huang2019}, 2019}  
\\  \hline  \hline
\textbf{Workflow Type} &
Scientific  & Scientific  & Scientific  & Business  & Business  
\\  \hline


\textbf{Multi-Tenancy} &
No & No & No & No & Yes 
\\  \hline

 \textbf{Workflow Targets} &
Intermediate Data &
  Intermediate Data, Task &
Intermediate Data &
Intermediate Data &
 Intermediate Data, Task 
\\  \hline

 \textbf{Covered Security Requirements} &
Data Integrity, Data Confidentiality & 
Data/Task Integrity, Data Confidentiality   &
Data Integrity &
Data Confidentiality, Data Integrity, Authentication &
Data Confidentiality, Task Confidentiality
\\  \hline

 \textbf{Considered Attackers} &
  Providers,  External  Attackers &
 Providers,  External Attackers &
 External  Attackers & 
 External  Attackers & 
Tenants 
 \\  \hline

 \textbf{Covered Attacks Categories} &
Network-based,   VM-based &  
Network-based, VM-based  & 
Application-based &
 Network-based, Application-based  &
Application-based 
\\  \hline

 \textbf{Covered Phases of the Workflow Lifecycle} &
  Execution, Monitoring  &
 Execution, Monitoring  &
  Execution, Monitoring  &
Modeling, Deployment &
  Execution, Monitoring  
\\  \hline

\bottomrule 
\end{tabulary}\par 
\end{table*}

A framework of a “mimic cloud workflow execution system”  is proposed in~\cite{Wang2018} featuring three strategies: heterogeneity (diversification of physical servers, hypervisors, and operating systems), redundancy (Lagged Decision Mechanism), and dynamics (switching workflow execution environment). This system only covers the execution and monitoring phases of the workflow life cycle and cannot carry out adaptation of the process instances to react to security violations. 

\cite{Mofrad2019} developed a secure big data workflow management which they called SecDATAVIEW, based on DATAVIEW \cite{Kashlev2014}. This system leverages the hardware-assisted trusted execution environments (TEEs) such as Intel Software Guard eXtensions (SGX) and AMD Secure Encrypted Virtualization (SEV) to protect the execution of big data workflows and the data used by them. They also proposed a secure architecture and the WCPAC (Workflow Code Provisioning and Communication) protocol for securing the execution of workflow tasks in remote worker nodes. This system is vulnerable to attacks like network traffic analysis, denial-of-service, side-channel attacks, and fault injections. Furthermore, it only protects workflows from possible attacks during execution, and if an attack occurs, it terminates the workflow execution. In other words, there is no alternative way to adapt the workflows to the detected violation in this system.

\cite{Kim2015} extended the Kepler provenance module and added the Security Analysis Package (SAP) to it in order to analyze provenance information in the security context using three security properties: input validation, remote access validation, and data integrity. This module can only detect some of the Application-based attacks and does not offer any way of  reacting to security violations during workflow execution. Besides, it does not consider providers as malicious actors.

\cite{Lins2016} proposed a system named BPA-Sec4Cloud, which aims to provide a "holistic and integrated cloud-based solution" to address the automation of security-aware business processes from modelling to their deployment. The system does not cover the monitoring, analysis, and adaptation phases of the lifecycle. Similar to \cite{Kim2015}, the providers are not considered as malicious actors.

\cite{Huang2019} presented a cloud workflow engine based on an extension of jBPM4~\cite{jBPM} that can support privacy protection between different tenant workflow instances in the cloud workflow systems. This system considers only malicious tenants as possible attackers and leaves out of scope attackers like service providers and users. Similarly, the solution cannot detect security violations in workflows and like others, does not allow for adaptations in the workflows as reaction to such violations.

We compare these WfMSs from different perspectives in Table \ref{table1} which shows that there is no WfMS that can protect all of the mentioned potential targets from all different types of actors (see Section \ref{sec:requirements}).

    \section{System Architecture}
\label{sec:architecture}

\begin{figure*}[t]
\setlength{\belowcaptionskip}{-12pt}
  \includegraphics[width=\textwidth]{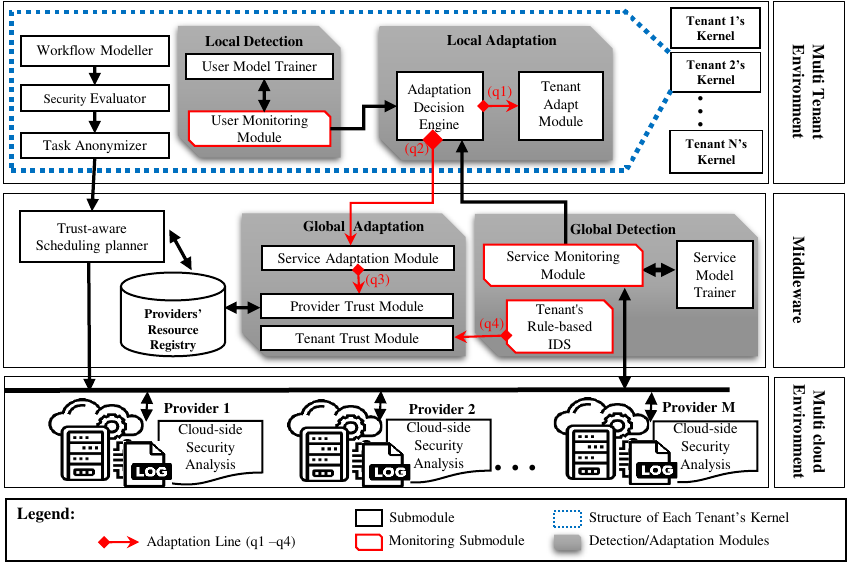}
  \caption{The architecture of SecFlow}
  \label{fig:Architecture}
\end{figure*}

This section introduces the architecture of our security-aware WfMS, \textit{SecFlow}, which is specifically designed to protect workflows from various security violations throughout their whole lifecycle. 
Figure \ref{fig:Architecture} provides a detailed view of the proposed architecture, highlighting its key modules such as the Tenant's Kernel, the Middleware, and the multi-cloud environment. We assume that the tenants' resources are cleanly isolated from each other and may be on the same cloud node; we also assume that the middleware is a logically centralized component that can be hosted by a third party. This deployment option we selected for our work offers the following benefits:
(a) It separates workflow instances of different tenants at runtime, meeting their specific functional and non-functional requirements \cite{Huang2019}, within isolated environments (i.e. the \textit{Tenant's Kernel}). This model also limits the amount of information the engine possesses about individual tenants. 
(b) It simplifies the cloud infrastructure for tenants. This is achieved by designing a logically centralized component -- the \textit{Middleware}, which facilitates informed decision-making for all tenants. The middleware component can be designed so as to be able to integrate with other middlewares, e.g. such that are used as communication backbones or service-oriented middlewares.

Other options of deploying \textit{SecFlow} are also possible but not in the scope of this work. 

\begin{figure}[bth!]

\includegraphics[width=\columnwidth]{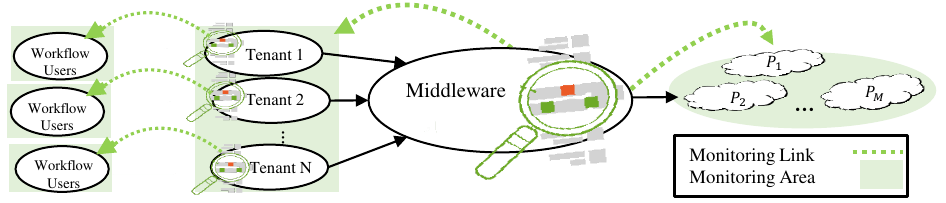}
\caption{General Overview of the Monitoring Procedure }
\label{fig:Architecture_simple}
\end{figure}

To meet the requirements of a security-aware WfMS, our work focuses on implementing a comprehensive monitoring procedure to detect potential attacks in the considered deployment model. Figure \ref{fig:Architecture_simple} provides a basic overview of this monitoring procedure, illustrating the locations of monitoring modules and their areas of responsibility. This monitoring approach aims to preserve privacy, safeguard sensitive information, and provide the capability to detect all possible attacks. In this procedure, tenants play an active role in monitoring their users. Similarly, the Middleware component supervises the behavior of both the Clouds and the tenants, using behavioral patterns learned from both sides. 

In the next sub-sections, we will describe the functionalities of the components of the \textit{SecFlow} architecture (excluding the multi-cloud environment).

 \subsection{Multi-tenant Environment}
\label{subsec:Tenant’s Kernel}
 Based on our assumption that each tenant has a cleanly separated version of the WfMS, the \textbf{Tenant's Kernel} provides an isolated environment for each tenant where they can model, analyze, execute, monitor, and adapt workflows based on their requirements and strategies. This ensures that the decisions and data of one tenant are protected from other tenants. 
 It consists of the following five sub-modules. 

 1)\textbf{\textit{Workflow Modeller:}} The Workflow Modeller module enables tenants to model their workflows in a secure and efficient way. During the modelling process, the security requirements (CIA) of each task are defined. 
 
 2) \textbf{\textit{Security Evaluator:}} This module is responsible for defining the possible adaptation actions for each task in the workflow. It involves establishing a set of adaptation strategies, such as skipping, re-working, re-sequencing, and re-configuration, of each task in the workflow instances that determine which actions are feasible for each task in the event of a security violation. Furthermore, it assesses the potential impact of each adaptation action on the overall value of the workflow. To illustrate this, let's consider the scenario of skipping a certain task within a workflow to mitigate the impact of a specific detected violation. Some tasks may be less critical, and skipping them may have minimal impact on the overall value of the workflow. However, for other tasks, such as authentication tasks, skipping is not a viable option in case of a violation, as they are crucial for maintaining the security of the workflow. 
 By incorporating task-specific adaptations into the security evaluator module (after the modeling phase), it becomes possible to estimate the cost of each adaptation action in terms of execution time and select the optimal adaptation strategy that has the lowest cost and maximal value. 
 
 3) \textbf{\textit{Task Anonymizer:}}
This module employs obfuscation techniques to securely handle sensitive data and removes unnecessary information for task processing. The obfuscated information is retained in intermediate results for future tasks, as required. Tenants within this module utilize client-side obfuscation techniques, such as data removal, noise injection, and data splitting, along with conflict detection methods to address conflicts between data-minimization and security requirements \cite{ Ramadan2020}. 

4) \textbf{\textit{Tenant's Detection module (Local Detection):}} Each tenant assumes the responsibility of training a machine learning model to detect malicious behavior among its users and monitors them based on this model. This approach not only ensures the security of each tenant's data associated with their users but also allows for customization based on the individual preferences of each tenant.  

5) \textbf{\textit{Tenant's Adaptation module (Local Adaptation):}} This module is responsible for selecting suitable adaptation actions according to the tenant's preferences and the run-time monitoring information, and performing these actions at the tenant level. 
The module comprises three sub-modules: a) \textit{Adaptation Decision Engine} to assess the cost of potential adaptation actions and prioritize those with the lowest impact on the system. It considers factors like price, time, mitigation impact, and overall value to the workflow in response to the detected attacks. Dependencies between tasks are also considered to prevent the propagation of violations so that the subsequent tasks are appropriately adapted. 
Tenant or middleware level actions are invoked based on the nature of each adaptation (e.g., \textbf{q1} and \textbf{q2} in Figure \ref{fig:Architecture}). b) \textit{Tenant Adapt Module} designed to perform the adaptations selected by the decision engine while still allowing for customized adjustments (e.g., skipping tasks, re-sequencing processes, introducing new tasks to mitigate the impact) per violation. 
The submodule can also respond to identified instances of malicious user behavior through appropriate adaptation measures. These measures may include lowering the user's trust level or imposing restrictions on their access to specific tasks. 
\vspace{-\baselineskip}

\subsection{Middleware}

This module is essential for ensuring the efficient management and scheduling of tasks for all tenants' kernels in the cloud environment. It monitors the behavior of the tenants and the cloud environment in order to detect any malicious activities that could potentially compromise the security of the workflow, and take appropriate actions based on the specific requirements of the submitted tasks. 
The module consists of three key components:
  
1) \textbf{\textit{Trust-aware Scheduling planner:}} The primary function of this module is to efficiently schedule workflows and allocate appropriate cloud resources for each task, taking into consideration the specific requirements of individual tenants, such as cost, time, and security.
The module integrates the regularly updated trustworthiness information of the providers received from the Provider Trust module into the scheduling process. It is also responsible for anonymizing the tenants' task specifications before transmitting them to the cloud for execution. This process is similar to the Task Anonymizer module, which operates on the tenant side.

2) \textbf{\textit{Global Monitoring and Detection: }} This module is responsible for real-time monitoring of cloud behavior by analyzing the cloud log file and the network traffic data. It includes two main modules:
a) \textit{Service Model Trainer:} 
The main purpose of this submodule is to train a robust machine learning model that can detect any malicious behavior in the cloud by analyzing the real-time network traffic data and cloud log files. The model is trained by using various parameters such as protocol type,  duration, and number of packets from the network traffic data sets. The cloud log file is also analyzed to extract information on CPU utilization, bandwidth consumption, and RAM utilization. 
b) \textit{Service monitoring:} The Service Monitoring submodule uses the machine learning model trained by the Service Model Trainer module to detect any malicious activity in the cloud services and providers and network attacks. It continuously analyzes the real-time network traffic data and cloud log file, comparing them with the expected behavior derived from the trained model. In case of  anomalies or suspicious activity, the module immediately raises an alert to the adaptation module of the corresponding tenant so that the appropriate measures are taken. 
c) \textit{Tenant's Rule-based Intrusion Detection System (IDS)} encompasses a collection of predefined rules in identifying potential attacks originating from the tenants by using submitted workload patterns and specific thresholds established for each tenant. 
The IDS can detect any suspicious or malicious activities exhibited by tenants during specific time intervals. When a tenant's behavior matches the predefined rules, indicating a potential attack, the IDS promptly triggers an alert (shown as \textbf{q4} in Figure \ref{fig:Architecture}).

3) \textbf{\textit{Global Adaptation: }} The Global Adaptation module is critical for ensuring efficient adaptation at the level of the middleware in response to the detected violations or security threats. It comprises three submodules described in detail below:

a) \textit{Service Adaptation Module:} This submodule enables the adaptation of services chosen by the Adaptation Decision Engine. Operating at the middleware level, this module is responsible for implementing the necessary actions to meet the evolving requirements of tenants. These actions may involve modifying the selected services within the same provider or exploring alternative services from different providers that align with the tenant's specific needs. Then, based on the detected violation, the trust score/level of the service and provider will be updated (shown as \textbf{q3} in Figure \ref{fig:Architecture}).

b) \textit{Provider Trust Module:} This submodule updates the trust level of the providers based on any detected violations or security risks. 
The updated trust values are utilized by the Trust-aware scheduling planner to schedule upcoming workflow instances, thus enhancing the overall security and efficiency of the system. Additionally, this module may modify the Provider Prediction model to improve the monitoring of malicious provider behavior with greater precision.

c) \textit{Tenant Trust Module:} Since the response to an attack in a tenant varies based on the attack's severity/impact, this submodule updates the tenant's trust level and takes corresponding actions, such as ignoring the alert (if trust falls below the defined threshold), isolating affected resources or activity blocking.

    \section{Evaluation}
\label{evaluation}
We implemented \textit{SecFlow} \footnote{Our code will be available soon at \textit{https://github.com/nafisesoezy/SecFlow}} by extending the jBPM (Java Business Process Management) \cite{jBPM} engine and integrating it with the Cloudsim Plus \cite{cloudsimplus} simulation tool. jBPM offers a pluggable architecture that allows for easy replacement of different module implementations. Additionally, the integration of the simulation framework Cloudsim Plus has allowed us to accurately model the complexities of a multi-cloud environment.

\subsection{Experimental Setting}
To evaluate our system,  we utilized three distinct categories of process models: Small (3-10 tasks), Medium (10-50 tasks), and Large (50-100 tasks). Our scenario assumed the availability of 5 cloud providers, each offering 3 different services for the service tasks. The specifications of these services fell within the following ranges: Response time [1, 50], Cost [0.1, 10], and confidentiality, integrity, and availability [0, 1]. 

Table \ref{table:Adaptation Types} provides an overview of the relative price and time associated with each adaptation type compared to the original task's response time ($R$) and price ($P$). The weights ($W$) are determined based on the workflow requirements provided. We utilize this table as a reference to determine the appropriate actions for each attack type (mitigated attackType) and its mitigation. 
To identify the optimal choice with minimal cost, we use Equation \ref{Equation:AdaptationCost} for computing the associated cost of each potential adaptation action. This equation factors in the price, time, and risk mitigation score specific to each adaptation action, while incorporating weights assigned by the tenant to prioritize their preferences.
\begin{align}
\begin{split}
AdaptationCost(aa,t) & = W_{price} \cdot  (AdaptPrice(aa) + PriceOverhead(aa,t)) + \\
& \quad  W_{time} \cdot  (AdaptTime(aa) + TimeOverhead(aa,t)) - \\
& \quad W_{Security} \cdot MitigationScore(aa,t)
\end{split}
\label{Equation:AdaptationCost}
\end{align}

In Equation \ref{Equation:AdaptationCost}, $AdaptPrice$, $AdaptTime$, and $MitigationScore$ represent the price, time, and risk mitigation score of the adaptation action $aa$, respectively. Additionally, $PriceOverhead$ and $TimeOverhead$ represent the adaptation price and time overhead specific to the adaptation action $aa$ for a given task $t$.

The calculation of $MitigationScore(aa)$ follows Equation \ref{Equation:MitigationScore}. It considers the security requirements of task $t$ (represented by $obj_t$), the impact of the detected attack $a_i$ on the CIA aspects (represented by $obj_{a_i}$), and the mitigation impact of the adaptation action on each aspect (represented by $obj_{aa}$).
\begin{align}
 MitigationScore(aa,t,a_i)= \sum_{obj \in \{C,I,A\}} (1 - obj_t \cdot obj_{a_i})*obj_{aa}
\label{Equation:MitigationScore}
\end{align}

In our experiments, we considered each adaptation action's mitigation impact (Table \ref{table:Adaptation Types}) and each attack's impact (Table \ref{table:Security Impacts}) on the CIA.

\begin{table*}[h!]
\fontsize{8pt}{10pt}\selectfont
\caption{{Cost of Different Adaptation Types} }
\label{table:Adaptation Types}
\def\arraystretch{1}
\ignorespaces 
\newcolumntype{L}[1]{>{\raggedright\let\newline\\\arraybackslash\hspace{0pt}}m{#1}}

\begin{tabulary}{\linewidth}{ 
L{2.2cm}L{1.8cm}L{1.7cm}L{2cm}L{2cm}L{1.8cm}}

 \textbf{AdaptType} &
\textbf{Late}  & 
\textbf{Skip}  &
\textbf{ReExecute} &
\textbf{Redundancy}&
\textbf{Reconfig}  
\\  \hline  \hline

\textbf{Time} &
$T*T_{Late}$  & $0$ & $T_{BackupSrc}$ & $T_{BackupSrc}$ & $T*T_{reconfig}$
\\  \hline

\textbf{Price} &
$P$  & $0$ & $P_{BackupSrc}$ & $P+P_{BackupSrc}$ & $P*P_{reconfig}$
\\  \hline

\textbf{Mitigation Impact(C,I,A)} &
$(0.7,0.6,0.8)$ & $(0.5,0.4,0.6)$ & $(0.8,0.9,0.7)$ & $(0.9,0.8,0.9)$ & $(0.6,0.7,0.5)$
\\  \hline

\textbf{AttackType Mitigated} &
DOS & Probe & DOS, Probe, U2R, R2L & DOS, U2R & DOS, Probe, U2R, R2L 
\\  \hline

\bottomrule 
\end{tabulary}\par

\end{table*}

\vspace{-0.5em}
\begin{table}
\caption{Security Impact of different attackTypes }
\label{table:Security Impacts}
\begin{tabular}{p{3cm} p{2.2cm} p{2.2cm} p{2.2cm} p{2.3cm}}
  \textbf{Attack type} &
 \textbf{DoS} &
\textbf{Probe}  & 
\textbf{U2R}  &
\textbf{R2L}  
\\  \hline
\bottomrule 
 \textbf{Impact on C,I,A} &
0.56,0.56,0.56	&      0.22, 0.22, 0	&     0.56, 0.22, 0.22 	&     0.56, 0.56, 0.22
\\  \hline
\end{tabular}\par 
\end{table}
\vspace{-0.5em}
\subsection{Main Results}
\begin{figure}[h]
\label{fig:evaluation}
  \includegraphics[width=\textwidth]{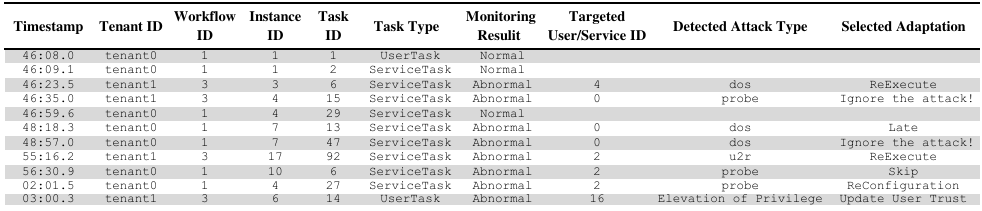}
  \caption{Logfile Snippet in SecFlow}
  \label{result-example}
\end{figure}

Figure \ref{result-example} presents a snippet from the system's logfile, providing insights into the activities of two tenants. At timestamp 46:08, the logfile entry reveals that $tenant0$'s $userTask1$ exhibits no indications of malicious behavior, as verified by the conducted user monitoring. Additionally, the logfile captures an occurrence of a violation within $tenant1$'s $serviceTask_6$, associated with the utilization of $service4$ from the available multi-cloud services. The detected attack type is identified as a Denial of Service (DoS) attack. In response to this threat, the architecture selects the adaptation strategy of $Reexecute$. This excerpt from the logfile showcases the architecture's capability to dynamically detect attacks originating from diverse entities. It showcases the architecture's adaptive nature, as it seamlessly adjusts its response strategy according to the type, severity, and characteristics of the detected attack, as well as the specific task in which the attack occurs. 

\begin{figure}[h]
\label{fig:evaluation}
  \includegraphics[width=\textwidth]{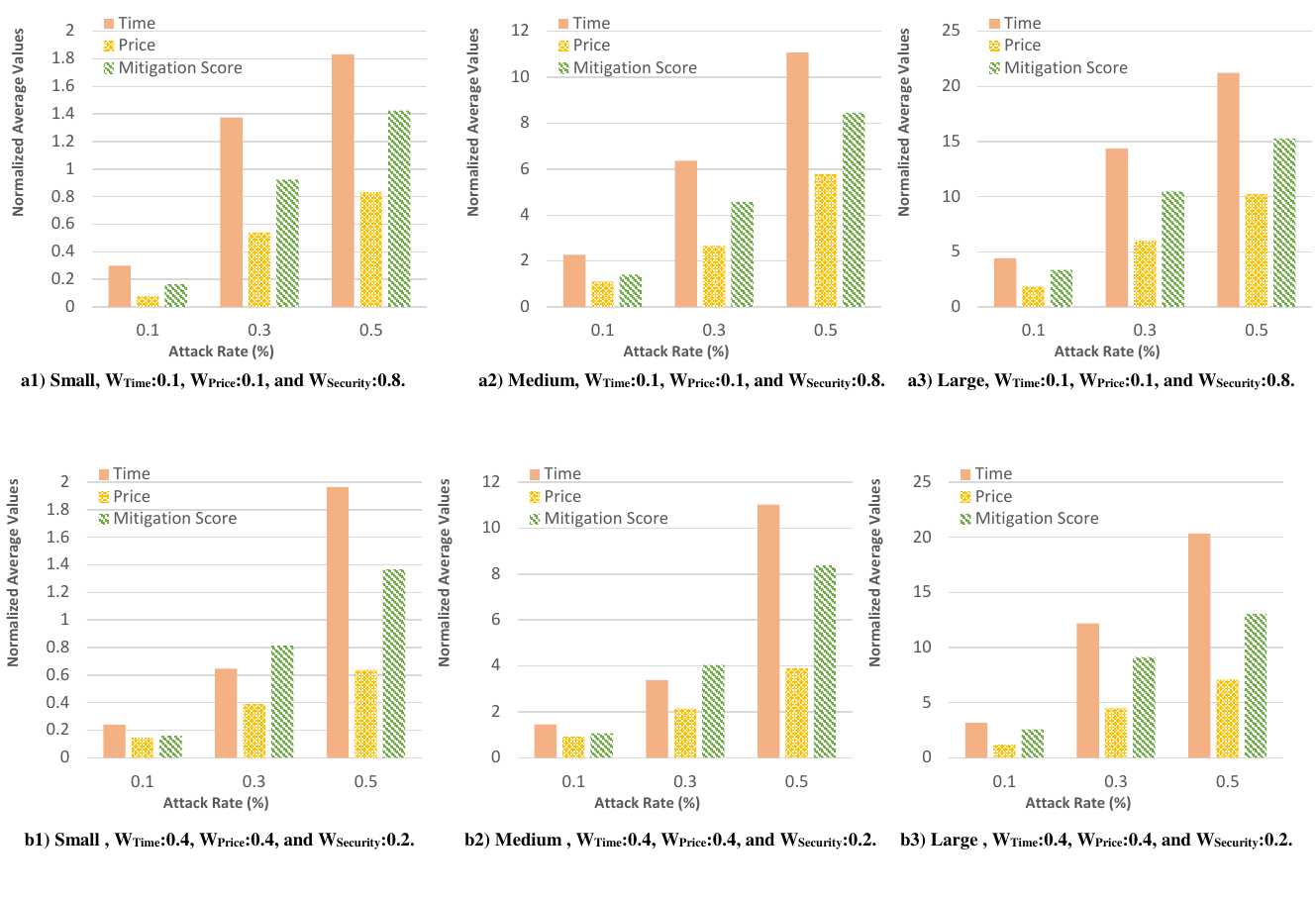}
  \caption{Normalized Average Time, Price, and Mitigation Score for Different Process Sizes with Varying Weights}
  \label{fig:finalResult}
\end{figure}

The results of our study are presented in Figure \ref{fig:finalResult}, which shows figures of the normalized average time, price, and mitigation score. These metrics were evaluated across 100 executions of three process categories (small, medium, and large) at varying attack rates. 

In Figure \ref{fig:finalResult}.a1, .a2, and .a3, we assigned weights of 0.1, 0.1, and 0.8 to time, price, and security respectively. Similarly, in Figure \ref{fig:finalResult}.b1, .b2, and .b3, the weights for time, price, and security were set as 0.4, 0.4, and 0.2, respectively.

Comparing these two sets of figures (Figure \ref{fig:finalResult}), we observe that the adaptation actions in b1, b2, and b3 have a shorter time and lower price compared to a1, a2, and a3. This reflects the tenant's higher prioritization of time and price in b1, b2, and b3. However, the mitigation scores in b1, b2, and b3 are lower than that of a1, a2, and a3, indicating a lower emphasis on the selected adaptation actions' mitigation effectiveness.

The findings highlight the effectiveness of the Adaptation Decision Engine Module in enabling tenants to tailor their adaptation strategies to meet their unique needs. By considering factors such as time, cost, and the mitigation score associated with various adaptation actions, tenants can strike a well-balanced approach that aligns with their requirements.

    \section{Conclusion}
\label{conclusion}

In this paper, we introduced SecFlow, a security-aware architecture designed for WfMS in a multi-cloud environment. Unlike previous studies, SecFlow comprehensively addresses security and privacy concerns throughout the entire workflow lifecycle, with particular emphasis on the detection and reaction to violations that are positioned in the adaptation phases of workflows. By considering threats from various parties, SecFlow provides an extensive monitoring functionality of malicious behavior at different levels (e.g., tenant and middleware) and detects abnormal activities. By leveraging the collected monitoring information, many adaptations become possible for safeguarding user privacy and/or tenant confidentiality.
The proposed architecture was implemented by extending the jBPM engine and integrating it with the Cloudsim Plus simulation tool. 
Experimental results demonstrate that SecFlow dynamically detects and responds to attacks while exhibiting good performance in terms of time, price, and mitigation score across workflows of different sizes.

As future work, we plan to extend the system's functionality to incorporate adaptive learning from past reactions and adaptations to violations. This enhancement will improve the overall effectiveness and responsiveness of SecFlow when securing and managing workflows in multi-cloud environments.

\section{Acknowledgments}

This work is partially funded by the HORIZON-KDT-JU-2022-1-IA project 101112089 AIMS5.0. The authors thank Dimka Karastoyanova for the input and contribution in most phases of this work.

\printbibliography
\end{document}